\documentclass[doublespace]{pasj01}

\Received{2015/03/03}
\Accepted{2015/10/31}

\usepackage{times}

\title{{Large-scale CO ({\bf {\it J}} = 4--3) Mapping toward the Orion-A Giant Molecular Cloud}}

\author{Shun \textsc{Ishii},\altaffilmark{1}\thanks{Present address: Institute of Astronomy, The University of Tokyo, 2-21-1 Osawa, Mitaka, Tokyo 181--0015} 
Masumichi \textsc{Seta},\altaffilmark{1}\thanks{Present address: School of Science and Technology, Kwansei Gakuin University, 2-1 Gakuen, Sanda, Hyogo 669-1337}
Makoto \textsc{Nagai},\altaffilmark{1} Yusuke \textsc{Miyamoto},\altaffilmark{1,2} Naomasa \textsc{Nakai},\altaffilmark{1} Taketo \textsc{Nagasaki},\altaffilmark{1,3} Hitoshi \textsc{Arai},\altaffilmark{1,2} Hiroaki \textsc{Imada},\altaffilmark{1} Naoki \textsc{Miyagawa},\altaffilmark{1} Hiroyuki \textsc{Maezawa},\altaffilmark{4} Hideki \textsc{Maehashi},\altaffilmark{1} Leonardo \textsc{Bronfman},\altaffilmark{5} and Ricardo\textsc{Finger}\altaffilmark{5}
}
\altaffiltext{1}{Division of Physics, Faculty of Pure and Applied Sciences, University of Tsukuba, 1-1-1 Ten-nodai, Tsukuba, Ibaraki 305-8571}
\altaffiltext{2}{Nobeyama Radio Observatory, NAOJ, 462-2 Nobeyama, Minamimaki, Nagano 384-1305}
\altaffiltext{3}{High Energy Accelerator Research Organization (KEK), 1-1 Oho, Tsukuba, Ibaraki 305-0801}
\altaffiltext{4}{The School of Science, Osaka Prefecture University, Osaka 599-8531}
\altaffiltext{5}{Departamento de Astronom\'{\i}a, Universidad de Chile, Casilla 36-D, Santiago, Chile}
\email{sishii@ioa.s.u-tokyo.ac.jp}
\begin{document}

\maketitle
\begin{abstract}
\vspace{-2mm}
We have mapped the Orion-A Giant Molecular Cloud in the CO (J = 4--3) line with the Tsukuba 30-cm  submillimeter telescope.
The map covered a 7.125 deg$\mathsf{^2}$ area with a $\mathsf{9'}$ resolution, including main components of the cloud such as Orion Nebula, OMC-2/3, and L1641-N. 
The most intense emission was detected toward the Orion KL region. 
The integrated intensity ratio between CO (J = 4--3) and CO (J = 1--0) was derived using data from the Columbia-Univ. de Chile CO survey, which was carried out with a comparable angular resolution. 
The ratio was $\mathsf{r_{4-3/1-0} \sim 0.2}$ in the southern region of the cloud and $\mathsf{0.4-0.8}$ at star forming regions.
We found a trend that the ratio shows higher value at edges of the cloud.
In particular the ratio at the north-eastern edge of the cloud at ($\mathsf{l}$, $\mathsf{b}$) $\mathsf{\approx}$ (\timeform{208.375D}, \timeform{-19.0D}) shows the specific highest value of 1.1.
The physical condition of the molecular gas in the cloud was estimated by non-LTE calculation.
The result indicates that the kinetic temperature has a gradient from north ($\mathsf{T_{kin}}$= 80 K) to south (20 K).
The estimation shows that the gas associated with the edge of the cloud is warm ($\mathsf{T_{kin}\sim60}$ K), dense ($\mathsf{n_{H_2}\sim10^4}$ cm$\mathsf{^{-3}}$), and optically thin, which may be explained by heating and sweeping of interstellar materials from OB clusters.
\end{abstract}

\section{Introduction}
Star forming activity in galaxies is closely related to the distribution and physical conditions of Giant Molecular Clouds (GMCs).
The large-scale surveys, which are represented by {Columbia-Univ.~de Chile} CO surveys, had investigated the global distribution of GMCs, structures of molecular gas both in space and velocity, and the total amount of the molecular gas in the Milky Way \citep{dame1987, bronfman1988, dame2001}.  In addition, the detailed structures, the size, and the mass of individual galactic regions and GMCs are also revealed by observations of $^{12}$CO ($J$=1--0), $^{12}$CO ($J$=2--1) and $^{13}$CO ($J$=1--0) lines, mostly with small and middle telescopes (e.g., \cite{sakamoto1994, dobashi1994, mizuno1995, oka1998}).
Nevertheless, the global distribution of physical conditions such as temperature, density and column density over the GMCs is still unclear because such observations have been carried out only in lower transition lines of CO. 

We made survey observations in the $^{12}$CO ($J$ = 4--3) emission line toward the Orion-A GMC to derive the physical properties of molecular gas over the GMC, with the same angular resolution of \timeform{9'} as those of the previous $^{12}$CO ($J$ = 1--0) and $^{12}$CO ($J$ = 2--1)  observations \citep{maddalena1986, wilson2005, sakamoto1994}.
This enables us to directly compare the intensities of the different CO transitions and thus to estimate the temperature and the column and volume densities of the gas. 
The Orion-A GMC is the most suitable site for studying massive star formation because of its close distance to the sun ($\sim 420$ pc by \cite{hirota2007, menten2007, kim2008}). Star formation is ongoing in the integral-shaped filament with the extent of $\sim 2^{\circ}$ ($\sim 15$ pc), along the north-south direction, which consists of several components such as OMC-1 to -5 \citep{johnstone1999, johnstone2006}.
\citet{maddalena1986} surveyed the Orion-Monoceros region in the $^{12}$CO ($J$ = 1--0) line with the 1.2-m telescope (\timeform{8'.7} of HPBW) as a part of Columbia-Univ.~de Chile CO survey.
They determined the spatial extent of three GMCs, Orion A, Orion B and Mon R2 (figure \ref{ori10}), where the mass was evaluated to be 1.0 $\times 10^5$, 0.8 $\times 10^5$, and 0.9 $\times 10^5 M_{\odot}$, respectively, comparing with the virial mass, the LTE mass, and the CO mass. 
\citet{wilson2005} re-observed the same region in the same line with uniform sampling.
\citet{sakamoto1994} mapped Orion-A and -B GMCs in the $^{12}$CO ($J$ = 2--1) line using the Tokyo-NRO 60-cm survey telescope\footnote{The telescope was renamed the AMANOGAWA telescope in 2010 after an upgrade to a new 2SB receiver in 2007 \citep{nakajima2007a, yoda2010}.}.
The $^{12}$CO ($J$ = 2--1)/$^{12}$CO ($J$ = 1--0) intensity ratio was almost unity on the main structure of the clouds, HII regions, reflection nebulae, and the western edge of the clouds. 
They estimated the gas density over the GMCs using the intensity ratio by the large velocity gradient method, and the densities of the main and peripheral regions were $\sim 3\times 10^3$ cm$^{-3}$ and $\sim 2\times 10^2$ cm$^{-3}$, respectively.
\citet{bally1987} surveyed the Orion-A GMC in the $^{13}$CO ($J$ = 1--0) line with the AT\&T Bell Laboratories 7-m telescope (\timeform{1'.8} of angular resolution).
The $^{13}$CO ($J$ = 1--0) map revealed that the overall structure of the cloud is filamentary.
This morphology indicates that the northern part of Orion A is compressed and supports massive star formation.
The southern part is however diffuse and exhibits chaotic spatial and velocity structure, supporting only intermediate- to low-mass star formation. 
The integral-shaped filament is about 0.5 pc wide, at least 13 pc long, and has a mass of $5\times10^3 M_{\odot}$ estimated from the $^{13}$CO ($J$ = 1--0) data.
\citet{bally1987}  explained these properties as resulting from the compression of the interstellar medium by a superbubble driven by the Orion OB association.  
\citet{nagahama1998} also mapped the Orion-A GMC in the $^{13}$CO ($J$ = 1--0) line and found 39 filamentary structures.
\citet{nishimura2015} mapped the same cloud in $^{12}$CO ($J$ = 2--1), $^{13}$CO ($J$=2--1), and C$^{18}$O ($J$ = 2--1) and derived temperature and density by comparing their data as well as $^{12}$CO ($J$ = 1--0), $^{13}$CO ($J$=1--0), and C$^{18}$O ($J$ = 1--0) data.
They found the temperature gradient along the cloud ridge. 
Mapping observations of $^{12}$CO ($J$ = 3--2) and [C$\;${\small\rmfamily I}]($^3$P$_1$--$^3$P$_0$) carried out by \citet{ikeda1999} with the Mt. Fuji submillimeter-wave telescope found that the distribution of [C$\;${\small\rmfamily I}]($^3$P$_1$--$^3$P$_0$) is quite similar to that of $^{13}$CO ($J$ = 1--0).

The GMC accompanies the Orion OB1 association, which is known as a typical stellar cluster \citep{brown1994}.
The association consists of several subgroups named OB1a-d from the northwest of the GMC.
The numbers of stars in subgroup a, b, c, and d are 53, 31, 34, and 3, respectively.  
Ages of the subgroups have been thought to be almost older from the northwest.
OB1a, the oldest subgroup is 11 Myr old.
OB1b and Ob1c were formed 1--5 Myr ago in the close generation, while \citet{brown1994} analyzed that the OB1c subgroup is older than the OB1b subgroup.
OB1d subgroup, located at the nearest to the Orion-A GMC and known as the Orion Nebula and NGC1976, was estimated less than 1 Myr old.  
The differences in ages and evolutional stages of the Orion-A OB1 subgroups and the GMC can be interpreted as the result of the sequential star formation, such as suggested by \citet{elmegreen1977}.  
The global star formation history of the Orion OB1 association and the GMC is therefore crucial to understand the formation of a stellar cluster and triggered star formation scenarios.

Several observations indicated that there are gradients in morphology, in velocity structure, and in intensity of emission lines over the Orion-A GMC.
The temperature-weighted velocity maps of the $^{12}$CO lines traced the substantial velocity gradient from the north ($V_{\mathrm{LSR}}=11$ km s$^{-1}$) to the south ($V_{\mathrm{LSR}}=3$ km s$^{-1}$) over the GMC \citep{wilson2005, shimajiri2011, nishimura2015}.
The origin of the velocity gradient has been attributed to rotation of the GMC \citep{kutner1977, maddalena1986} or to large scale expansion driven by the stellar winds of the Orion OB1 association \citep{bally1987}. 
\citet{wilson2005} also advocated the latter, because there were evidences of interaction between the Orion OB1 association and the Orion-A GMC. 
For example, there are sharp $^{12}$CO ($J$ = 1--0) intensity drop and velocity shifts with respect to the main component of the cloud at the edge of the cloud, which are attributed to expanding shells driven by an HII region, the Orion Nebula, excited by young massive stars in OB1c. 
The integrated intensity ratio of [C$\;${\small\rmfamily I}]($^3$P$_1$--$^3$P$_0$)/$^{12}$CO ($J$ = 3--2) also shows a gradient from north ($\sim$0.1) to south ($\sim$1.2) \citep{ikeda2002}. 
The gradient of the [C$\;${\small\rmfamily I}]($^3$P$_1$--$^3$P$_0$)/$^{12}$CO ($J$ = 3--2) ratio was interpreted as a consequence of the temperature gradient.
In addition, heating by the Orion OB1 association may be responsible for the global gradient.  
The diffuse interstellar material may have been swept up and accumulated by the OB1 association through the processes that made the temperature and intensity gradients in the Orion-A GMC. 
Some authors suggested that the triggered star formation is occurring by the interaction between the Orion OB1 association and the Orion-A GMC \citep{wilson2005, shimajiri2011}. 

In the next section, we describe the details of the observations. 
The results of the survey, the distribution and velocities over the GMC, are presented in section 3. 
In section 4, we analyze the physical properties of the molecular gas using the $^{12}$CO ($J$ = 4--3) data with other archival data and also discuss the triggered star formation in the GMC.
In the last section, we summarize the results of this paper.
Throughout this paper, we adopt 418 pc \citep{kim2008} as the distance of Orion KL, at which \timeform{1'} and \timeform{1''} correspond to 0.12 pc and $2.0 \times 10^{-3}$ pc, respectively. The velocity is used here expressed with respect to the local standard of rest (LSR) in the radio definition.

\section{Observations}

\begin{figure}[tbp]
\begin{center}
\includegraphics[width=80mm]{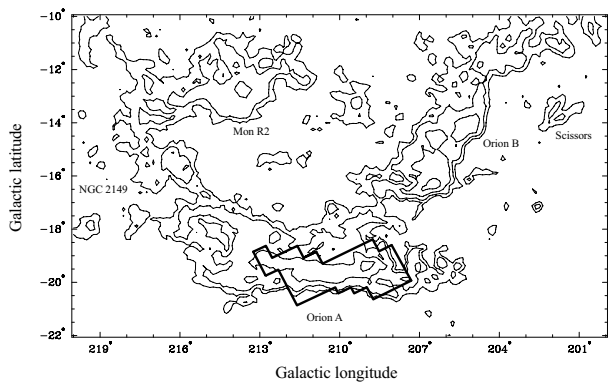}
\end{center}
\caption{The distribution of the Orion-Monoceros molecular complex traced by the integrated intensity of the $^{12}$CO ($J$ = 1--0) emission \citep{wilson2005}. The contour levels are 3, 9, 27, and 81 K km s$^{-1}$. The mapping region of the $^{12}$CO ($J$ = 4--3) line are indicated with thick lines.}
\label{ori10}
\end{figure}

\begin{figure}[tbp]
\begin{center}
\includegraphics[width=80mm]{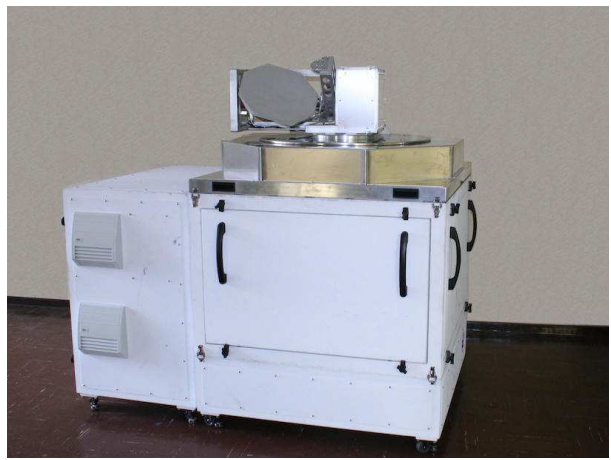}
\end{center}
\caption{Photograph of the Tsukuba 30-cm submillimeter telescope which is of the offset Cassegrain design.}
\label{30cm}
\end{figure}

\begin{table*}
\tbl{Summary of $^{12}$CO ($J$=4--3) observations with the Tsukuba 30-cm telescope}{%
\begin{tabular}{ll}
 \hline
     Telescope	&  \\ \hline
     Antenna diameter &	30 cm	\\
     Angular resolution (HPBW)	& \timeform{9'.4} $\pm$ \timeform{0'.4}	\\
     Main beam efficiency & 0.87 $\pm$ 0.10 \\
     Receiver noise temperature &  900 K \\
     Sideband rejection ratio &  $21 \pm 4$ dB \\
     Bandwidth of spectrometer	& 40 -- 960 MHz ($\pm$ 300 km s$^{-1}$ at 461 GHz)  \\
     Resolution of spectrometer 	& 61 kHz (0.04 km s$^{-1}$ at 461 GHz) \\
     \hline 
     Site 		&  \\ \hline
     Place		& Parinacota, Chile\\
     Location	&  \timeform{18D12'} S,  \timeform{69D16'} W \\
     Altitude 	& 4400 m \\
     \hline 
     Observations			&  \\ \hline
     Date 					& 2011 September 	15 -- October 17\\
     Rest frequency 	of $^{12}$CO ($J$=4--3) 	& 461.04077 GHz \\
     System noise temperature &  2500 -- 4000 K \\
     Pointing  accuracy		& $<$ \timeform{1'.0} \\ 
     Observing area			& 7.125 degree$^2$ \\
     Numbers of positions		& 685 \\ 
     Grid spacing			& \timeform{7'.5} (\timeform{3'.25} around Orion KL) \\
     Reference position		& ($\alpha_{2000}$, $\delta_{2000}$) $=$ (\timeform{05h32m12.7s}, \timeform{-6D27'53''}) \\
     Noise level in $T_{\mathrm{mb}}$ scale		& 0.5 K  with smoothing velocity of  0.6 km s$^{-1}$\\
 \hline
\end{tabular}}\label{table}
\end{table*}

The observations of the $^{12}$CO ($J$ = 4--3) were made with the Tsukuba 30-cm submillimeter telescope of the University of Tsukuba during 2011 September and October.
The 30-cm telescope (figure \ref{30cm}) is a transportable submillimeter telescope developed for a Galactic Plane survey in the CO ($J$ = 4--3) line at 461.04077 GHz and [C$\;${\small\rmfamily I}] ($^3$P$_1$--$^3$P$_0$) line at 492.16065 GHz from the Dome Fuji station on the Antarctic plateau \citep{seta2012, ishii2013}.
We operated the telescope at a 4400-m site above sea level at Parinacota in northern Chile for test observations and carried out the present survey of Orion-A GMC. 
The site is located at \timeform{18D12'} S and \timeform{69D16'} W.
The telescope has a 30-cm diameter of the main reflector.
The half-power beam width of the telescope was HPBW=\timeform{9'.4} $\pm$\timeform{0'.4} at 461 GHz, which was measured by scanning the Sun and corresponds to 1.1 pc at the distance of Orion KL, 418 pc. 
The beam size is almost equal to those of the Columbia-CfA and Univ. de Chile 1.2-m telescopes at the frequency of CO ($J$=1--0) \citep{dame2001} and the AMANOGAWA telescope at the CO ($J$=2--1) frequency \citep{yoda2010}.
The forward coupling efficiency was estimated by observing the new moon to be $\eta_{\mathrm{Moon}} = 87$\%$ \pm10 $\% by assuming the brightness temperature of the new moon to be 110 K at 461 GHz \citep{linsky1973}. 
The sidelobe level of the beam of the 30-cm telescope was less than $-18$ dB relative to the peak of the main beam. 
In this condition we can regard the forward coupling efficiency as the main beam efficiency $\eta_{\mathrm{mb}}$ and the difference between $\eta_{\mathrm{Moon}}$ and $\eta_{\mathrm{mb}}$ is estimated to be $\lesssim 1\%$ \citep{yoda2010}. 
We adopt  $\eta_{\mathrm{mb}} \approx \eta_{\mathrm{Moon}} = 87$\%$ \pm 10$\% in this paper.

We used an SIS mixer receiver with noise temperature of $T_{\mathrm{rx}} \sim$ 900 K in a single sideband mode realized by a quasioptical Martin-Pupplet type filter in front of the feed horn \citep{manabe2003}. 
This little bit high receiver noise temperature results from higher physical temperature of the mixer and image signal terminating load due to limited cooling capability of a compact cryocooler.
The sideband rejection ratio was measured to be $21 \pm 4$ dB at 461 GHz.
The typical system noise temperature including the atmospheric effect was 2500--4000 K at the observing elevations.
The receiver backend was an FX type digital spectrometer with 16384 channels, with a total bandwidth and frequency resolution of 1 GHz and 61 kHz, corresponding to 650 km s$^{-1}$ and $0.04$ km s$^{-1}$, respectively at 461 GHz.
A band pass filter was inserted in front of the spectrometer to avoid the aliasing effect, so the effective bandwidth of the spectrometer was 40--960 MHz that covered 600 km s$^{-1}$ in velocity.
We observed $685$ positions in the region of $7.125$ deg$^2$ on the sky (figure \ref{ori10}).
All observations were carried out in position switching mode, with a selected reference position of ($\alpha_{2000}$, $\delta_{2000}$)$=$(\timeform{05h32m12.7s}, \timeform{-6D27'53.4''}) adopted from the CO ($J$ = 3--2) mapping with the Mt. Fuji submillimeter-wave telescope \citep{ikeda1999}. 
The observed positions were spaced every \timeform{7.5'} in the equatorial coordinate.
Additional data around Orion KL were obtained with a grid spacing of \timeform{3.75'}.

Pointing of the telescope was checked by observing bright stars using an optical CCD camera mounted on a stay of the subreflector.
In addition, we observed CO ($J$ = 4--3) of Orion KL several times a day with a nine point cross scan of \timeform{4'.5} spacing in order to check variations of the pointing and intensity scale.
Measured pointing accuracy was always better than \timeform{1'}, corresponding to about one tenth of the beam size.
The calibration of the line intensity was made using a black body at ambient temperature by the standard chopper-wheel method \citep{penzias1973, ulich1976} every 15 minutes or more often, yielding an antenna temperature, $T_{\mathrm{A}}^{*}$, corrected for the atmospheric and antenna ohmic losses.
Relative intensity variations of Orion KL were within $10$\% in rms.
In this paper, we use as the intensity scale the main-beam brightness temperature, defined by $T_{\mathrm{mb}}=T_{\mathrm{A}}^{*}/\eta_{\mathrm{mb}}$. 

We achieved an rms noise level of $\Delta T_{\mathrm{mb}} = 0.5$ K at a resolution of $0.6$ km s$^{-1}$ for the CO ($J$ = 4--3) line data after integrating typically 5 minutes per position.
We reduced the data using the NEWSTAR package, which is reduction software developed by the Nobeyama Radio Observatory. 
The reduction procedure included flagging out bad data, integrating the data, and removing a slope in the base line by least-square fitting of the first order polynomial. The details of the observations are summarized in Table \ref{table}.

\section{Results}

\begin{figure}[tbp]
\begin{center}
\includegraphics[width=80mm]{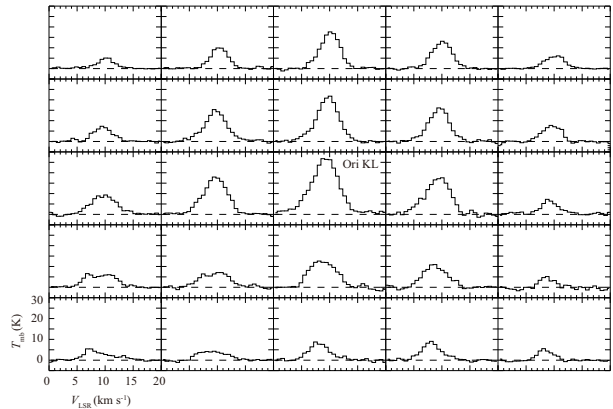}
\end{center}
\caption{Samples of the CO ($J$ = 4--3) spectra measured around Orion KL, parallel to the Right Ascension and Declination. For each spectrum, the abscissa is the LSR velocity ($V_{\mathrm{LSR}}=0 - 20 $ km s$^{-1}$) and the ordinate is the main-beam brightness temperature ($T_{\mathrm{mb}}$= -5 -- 30 K). The grid spacing between the spectra is $7'5$.
}
\label{oriKL43}
\end{figure}

\begin{figure*}[tbp]
\begin{center}
\includegraphics[width=150mm]{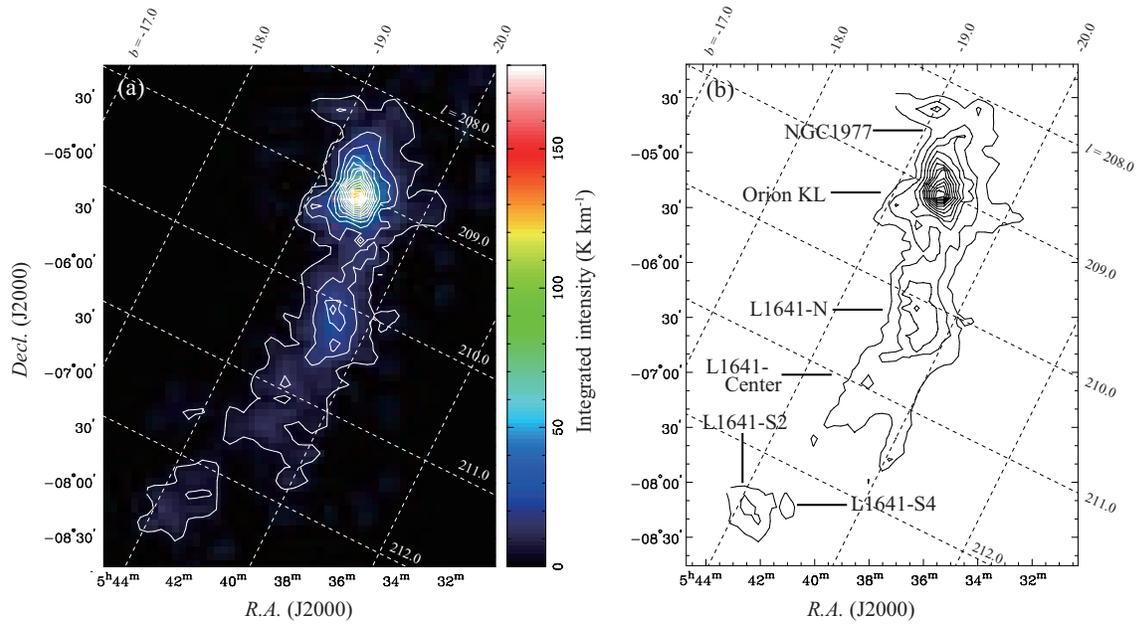}
\end{center}
\caption{($a$) Integrated intensity map of $^{12}$CO ($J$ = 4--3). The contour levels are from $I_{\mathrm{CO}}=5$ K km s$^{-1}$ to 165 K km s$^{-1}$ with an interval of 10 K km s$^{-1}$. ($b$) Peak main-brightness temperature map of $^{12}$CO ($J$ = 4--3). The contour levels are from $T_{\mathrm{mb}}=2$ K to 26 K with an interval of 2 K.}
\label{co43int}
\end{figure*}

\begin{figure*}[tbp]
\begin{center}
\includegraphics[width=150mm]{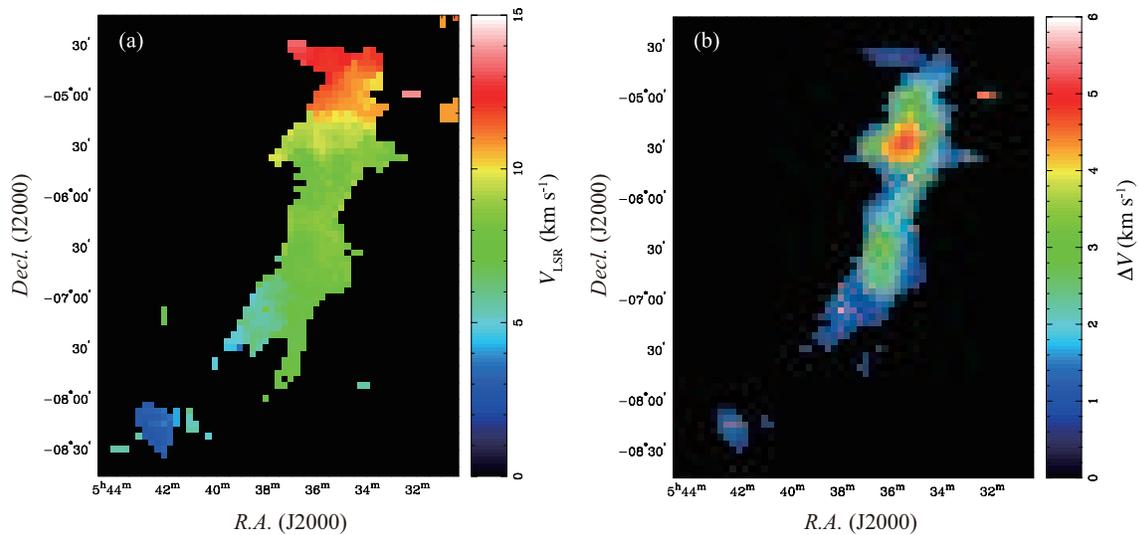}
\end{center}
\caption{Maps of spatial distribution of ($a$) temperature-weighted velocity and ($b$) velocity width (FWHM) of $^{12}$CO ($J$ = 4--3) in the region of $T_{\mathrm{mb}} >$ 2.0 K.}
\label{mom}
\end{figure*}

Figure \ref{oriKL43} shows samples of the CO ($J$ = 4--3) spectra measured around Orion KL.
Figure \ref{co43int}a shows the intensity map of CO ($J$ = 4--3) integrated over the velocity range of $V_{\mathrm{LSR}}=0-20$ km s$^{-1}$, $I_{\mathrm{CO}}=\int T_{\mathrm{mb}} dv$.
The map indicates an elongated cloud north to south with its length of  $\sim$30 pc (4.2 degrees) and width of $\sim$6 pc ($50'$).
The strongest emissions appears at the position of Orion KL with a peak temperature, $T_{\mathrm{mb}} = 26.8$ K, an integrated intensity of $I_{\mathrm{CO}}=168.4$ K km s$^{-1}$, a line width of $\Delta V = 5.4$ km s$^{-1}$, and a peak velocity of $V_{\mathrm{LSR}} = 9.5$ km s$^{-1}$.
The obtained intensity and peak velocity are compared with the previous observations by \citet{schulz1995} after correcting an effects due to the difference of the angular resolution of the telescopes and are consistent within the calibration error  (see details in \cite{ishii2013}). 

In the southern part of Orion A, bright and faint CO ($J$ = 4--3) regions are seen at L1641-N ($\alpha_{2000}$, $\delta_{2000}$ $=$ \timeform{05h36m19.0s}, \timeform{-6D22'13.3''}), L1641-S2 (\timeform{05h42m47.2s}, \timeform{-8D17'5.5''}), and L1641-S4 (\timeform{05h40m48.8s}, \timeform{-8D06'50.9''}) that are known as molecular outflows associated with star forming region(s) of low/intermediate-mass stars \citep{fukui1989}. 
Figure \ref{co43int}b shows the peak brightness temperature map. 
Another peak is also located at (\timeform{05h38m00s}, \timeform{-7D05'00''}), at the west of L1641-Center (\timeform{05h38m46.4s}, \timeform{-7D01'5.0''}) identified as well in CO ($J$ = 1--0).
In the north of Orion A, a peak near NGC 1977 at (\timeform{05h35m30s}, \timeform{-4D35'00''}) is manifested only in the peak temperature map of CO ($J$ = 4--3).
The CO emission decreases steeply in both the temperature and the integrated intensity maps, to the northern east Orion KL.
This is consistent with the results of the previous observations in $^{12}$CO ($J$ = 1--0) and $^{13}$CO ($J$ = 1--0) \citep{bally1987, wilson2005}. 
In addition, the east edge of L1641-N also shows steep change in the peak temperature map, with the emission from the gas component extending toward the south-west direction of L1641-N.

\begin{figure}[tbp]
\begin{center}
\includegraphics[width=80mm]{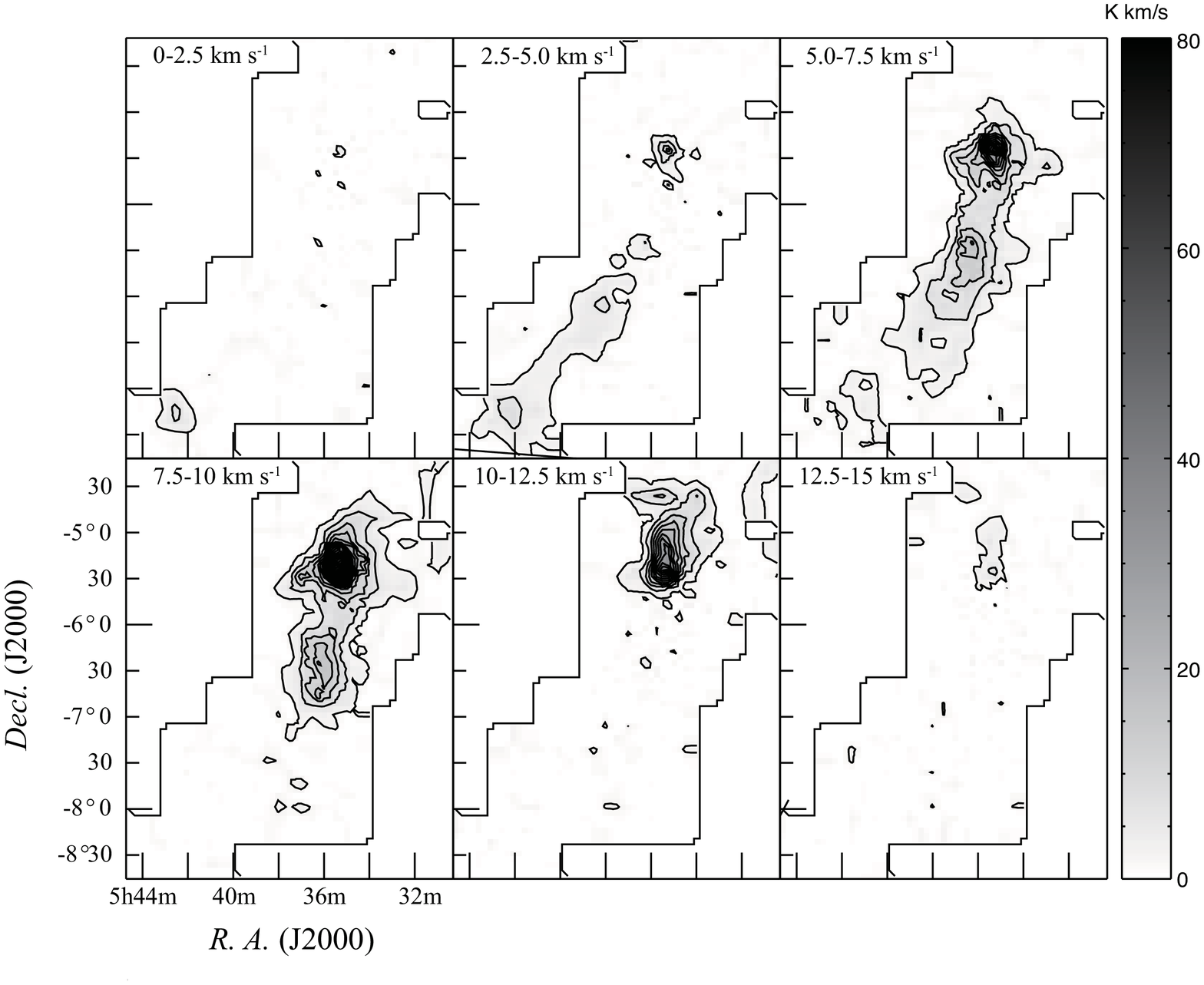}
\end{center}
\caption{Channel map of $^{12}$CO ($J$ = 4--3). The velocity range are indicated at the top left corner of each map. The contour level are from 2 K km s$^{-1}$ to 80 K km s$^{-1}$ with an interval of 4 K km s$^{-1}$.}
\label{CO43chmap}
\end{figure}

\begin{figure}[tbp]
\begin{center}
\includegraphics[width=80mm]{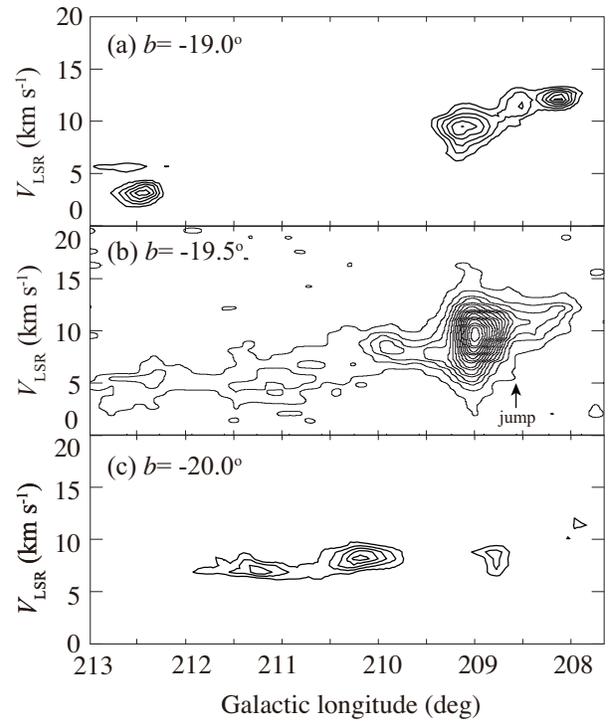}
\end{center}
\caption{Longitude-velocity diagram of $^{12}$CO ($J$ = 4--3) at ($a$) $b$ = \timeform{-19.0D}, ($b$) $b$ = \timeform{-19.5D}, and ($c$) $b$ = \timeform{-20.0D}. The contour levels are from $T_{\mathrm{mb}}=2$ K with an interval of 1 K.}
\label{pv}
\end{figure}

Figure \ref{mom}a shows the temperature-weighted mean velocity (1st moment map) defined by $\int T_{\mathrm{mb}} v dv/\int T_{\mathrm{mb}} dv$.
The CO ($J$ = 4--3) profiles show velocity components beyond 14 km s$^{-1}$ around HII regions, M42 (Orion KL) and NGC 1977 (e.g., figure 3), as seen in $^{12}$CO ($J$ = 1--0) but not in $^{13}$CO ($J$ = 1--0) nor in [C$\;${\small\rmfamily I}] \citep{bally1987, nagahama1998, ikeda2002}.
In the southern region, from Decl. $=$\timeform{-7D30'} to Decl. $=$\timeform{-6D45'}, the mean velocity at the east side of the clouds has 2--3 km s$^{-1}$ lower than that at the west side.
Figure \ref{mom}b shows a line width map calculated from the 2nd order moment map.
The large velocity width of  $\Delta V > 4$ km s$^{-1}$ is distributed in the region within $\sim$ \timeform{15'} of the Orion KL center.
The broad velocity width of 3 km s$^{-1}$ is also detected at the east side of the L1641-N and at the CO integrated intensity peak of L1641-S4. 
The broad CO emission at L1641-S4  corresponds to a component in channel maps of 0--2.5 and 2.5--5.0 km s$^{-1}$ (figure \ref{CO43chmap}).
The high velocity component on the north has a small velocity width of 1.7 km s$^{-1}$.

Figure \ref{pv} shows position-velocity diagrams along the galactic longitude at $b=$\timeform{-19.0D}, $b=$\timeform{-19.5D}, and $b=$\timeform{-20.0D}. 
It shows the velocity gradient from 2 to 14 km s$^{-1}$ along the center of the cloud as noticed in the the temperature-weighted velocity map. 
The CO ($J$ = 1--0) emission has a velocity jump at $l \sim$ \timeform{208.5D}  in the position-velocity diagram (figure. 5b in \cite{wilson2005}), where there is no low velocity component ($\leq 5$ kms$^{-1}$) beyond the jump point. This suggests that the low velocity molecular gas has been cleared at $l<\timeform{208.5}$. The same velocity jump was detected in the CO ($J$ = 4--3) at $l \approx \timeform{208.75}$ (figure \ref{pv}b).
It forms a CO edge at $l =$ \timeform{210.0D} and  there is no CO ($J$ = 4--3) at $V_{\mathrm{LSR}}< 8$ km s$^{-1}$ at $l <$ \timeform{208.5D}.
Figure \ref{CO43chmap} also represents this trend.
The filamentary structure has low ($\sim$6 km s$^{-1}$) and high velocity ($\sim$12 km s$^{-1}$) components at the west side of Orion KL.
L1641-S4 has wide velocities of 1.5 km s$^{-1}$ to 7 km s$^{-1}$.

\section{Discussion}

We derive the physical conditions of the molecular gas in the Orion-A GMC by a non-LTE analysis, combining our $^{12}$CO ($J$ = 4--3) data and archival data, $^{12}$CO ($J$ = 1--0) and $^{13}$CO ($J$ = 1--0).
$^{12}$CO ($J$ = 1--0) data are a part of the Columbia survey \citep{dame2001, wilson2005}. 
The survey mapped the region of Orion A with a spatial resolution of $8'.4$ and a grid spacing of \timeform{0.125D} in the galactic coordinates. 
The velocity resolution is 0.65 km s$^{-1}$,  the same as $^{12}$CO ($J$ = 4--3). 
The typical rms noise level is $\Delta T_{\mathrm{mb}} = 0.26$ K. 

For $^{13}$CO ($J$ = 1--0), the data taken with the Bell 7-m telescope \citep{bally1987} are adopted, where the beam size was HPBW= \timeform{1'.8}. 
The data consisted of 33000 positions in the region of 8 deg$^2$ with a grid spacing of $1'$. 
Spectra covered from $V_{\mathrm{LSR}}=1.0$ to 14 km s$^{-1}$ with the velocity resolution of 0.27 km s$^{-1}$. 
The noise level was $\Delta T_{\mathrm{mb}}=0.30$ K. 

\subsection{Integrated intensity maps}
The data grid of the $^{12}$CO ($J$ = 4--3) and $^{13}$CO ($J$ = 1--0) maps were converted from the equatorial coordinates to the galactic coordinates to be coincident with the $^{12}$CO ($J$ = 1--0) map, at the observing positions. 
We also convolved the $^{13}$CO ($J$ = 1--0) data with a gaussian function to yield the angular resolution of FWHM$=9'.4$, the same as that of the $^{12}$CO ($J$ = 4--3). 

We calculated the intensity of each line integrated from 1.0 km s$^{-1}$ to 14.0 km s$^{-1}$, $I_{^{12}\mathrm{CO (4-3)}}, I_{^{12}\mathrm{CO (1-0)}}$, and $I_{^{13}\mathrm{CO (1-0)}}$, for the analysis.
As mentioned above, a small fraction of the molecular gas associated with Orion KL has a velocity $V_{\mathrm{LSR}} > 14$ km s$^{-1}$, but the effect of the higher velocity gas to the present analysis is negligible.
Averages of the integrated intensities are $I_{^{12}\mathrm{CO (4-3)}}=14.6$, $I_{^{12}\mathrm{CO (1-0)}}=58.8$, and $I_{^{13}\mathrm{CO (1-0)}}=8.3$ K km s$^{-1}$ in the mapped region.
The rms uncertainty of the integrated intensity at each position is described as 
\begin{equation}
\Delta I = \Delta T \sqrt{\Delta V \Delta V_s},
\end{equation}
where $\Delta T$ is the rms noise, $\Delta V$ the velocity width of a spectrum, and $\Delta V_s$ the smoothed velocity resolution.
Typical values of $\Delta I$ for $I_{^{12}\mathrm{CO (4-3)}}, I_{^{12}\mathrm{CO (1-0)}}$, and $I_{^{13}\mathrm{CO (1-0)}}$ are 0.5, 0.9, 0.4 K km s$^{-1}$, respectively, for the average line width of the three lines, $\Delta V = 5.0$ km s$^{-1}$.

\begin{figure}[tbp]
\begin{center}
\includegraphics[width=80mm]{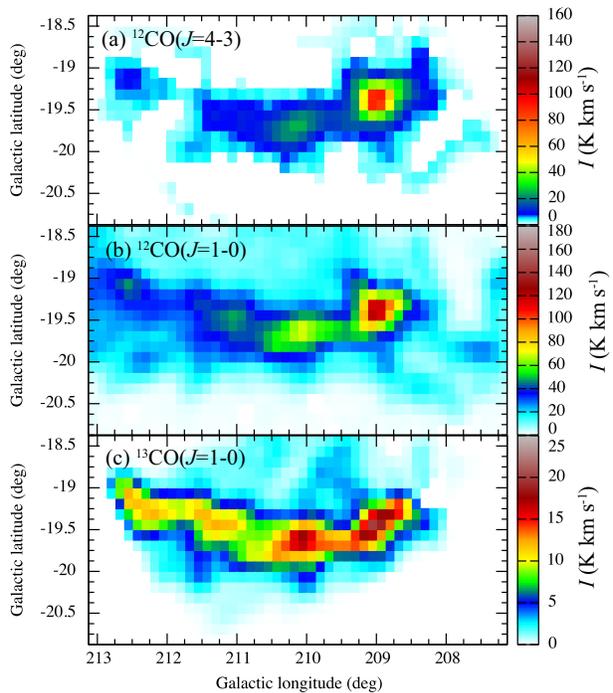}
\end{center}
\caption{Integrated intensity map of ($a$) $^{12}$CO ($J$ = 4--3), ($b$) $^{12}$CO ($J$ = 1--0), and ($c$) $^{13}$CO ($J$ = 1--0), where the data of $^{12}$CO ($J$ = 1--0) and $^{13}$CO ($J$ = 1--0) were adopted from \citet{wilson2005} and \citet{bally1987}.}
\label{Ico_gal}
\end{figure}

\begin{figure*}[tbp]
\begin{center}
\includegraphics[width=150mm]{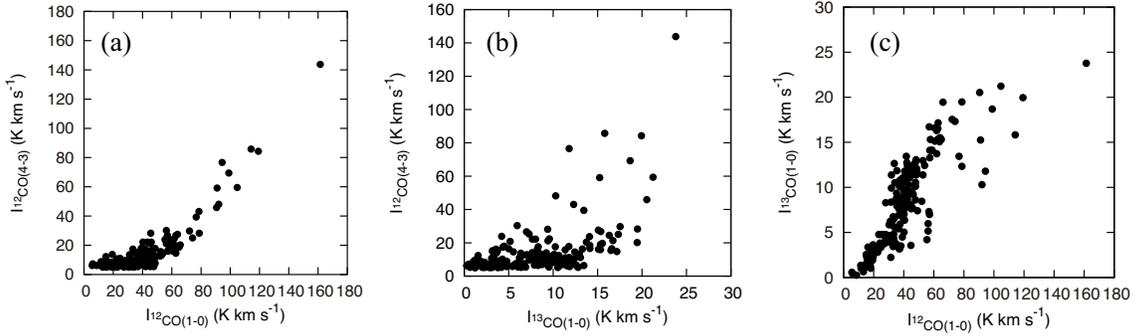}
\end{center}
\caption{Correlation of the integrated intensities between ($a$) $I_{^{12}\mathrm{CO (1-0)}}$ and $I_{^{12}\mathrm{CO (4-3)}}$, ($b$) $I_{^{13}\mathrm{CO (1-0)}}$ and $I_{^{12}\mathrm{CO (4-3)}}$, and ($c$) $I_{^{12}\mathrm{CO (1-0)}}$ and $I_{^{13}\mathrm{CO (1-0)}}$, where the data of $^{12}$CO (1-0) and $^{13}$CO (1-0) were adopted from \citet{wilson2005} and \citet{bally1987}.}
\label{correlation}
\end{figure*}

Figure \ref{Ico_gal} shows the integrated intensity map of the three lines.
The integrated intensity map of CO ($J$ = 4--3) emission exhibits smaller spatial distribution than those of CO ($J$ = 1--0) and CO ($J$ = 2--1) emissions \citep{wilson2005, sakamoto1994} despite similar noise levels.
Figure \ref{correlation} shows correlation plots between the integrated intensities of the three lines. 
Figure \ref{correlation}a indicates a good correlation between $I_{^{12}\mathrm{CO (4-3)}}$ and $I_{^{12}\mathrm{CO (1-0)}}$ at $I_{^{12}\mathrm{CO (1-0)}} > 60$ K km s$^{-1}$, but $I_{^{12}\mathrm{CO (4-3)}}$ is weak at $I_{^{12}\mathrm{CO (1-0)}}< 40$ K km s$^{-1}$.
CO molecules are not populated well for the density corresponding to $I_{^{12}\mathrm{CO (1-0)}} < 40$ K km s$^{-1}$. The correlation of $I_{^{12}\mathrm{CO (4-3)}}$ with $I_{^{13}\mathrm{CO (1-0)}}$ can be seen in figure \ref{correlation}b, but the dispersion of the scatter plot is larger than that with the dispersion of $I_{^{12}\mathrm{CO (1-0)}}$. The $I_{^{13}\mathrm{CO (1-0)}}$- $I_{^{12}\mathrm{CO (1-0)}}$ plot in figure \ref{correlation}c has comparable dispersion.
As the case with the $I_{^{12}\mathrm{CO (1-0)}}$ plot, $I_{^{12}\mathrm{CO (4-3)}}$ is weak for $I_{^{13}\mathrm{CO (1-0)}}< 15$ K km s$^{-1}$.

\subsection{Non-LTE analysis}

\subsubsection{Correlation between the intensities of CO ($J$=4-3) and lower CO transitions}

Intensity ratios can be used to probe the rough trend of the physical conditions of the molecular gas. 
We define beam-averaged intensity ratios between $^{12}$CO ($J$ = 4--3), $^{12}$CO ($J$ = 1--0), and $^{13}$CO ($J$ = 1--0),
\begin{eqnarray}
r_{4-3/1-0}=I_{^{12}\mathrm{CO (4-3)}} /I_{^{12}\mathrm{CO (1-0)}}, \\
r_{4-3/13}=I_{^{12}\mathrm{CO (4-3)}} /I_{^{13}\mathrm{CO (1-0)}}, \\
r_{13/12}=I_{^{13}\mathrm{CO (1-0)}} /I_{^{12}\mathrm{CO (1-0)}},
\end{eqnarray}
where the beam sizes for the lines are almost the same. Typical uncertainties of $r_{4-3/1-0}$, $r_{4-3/13}$, and $r_{13/12}$ are $\Delta r_{4-3/1-0}=0.02$, $\Delta r_{4-3/13}=0.03$, and $\Delta r_{13/12}=0.01$, respectively.

\begin{figure}[tbp]
\begin{center}
\includegraphics[width=80mm]{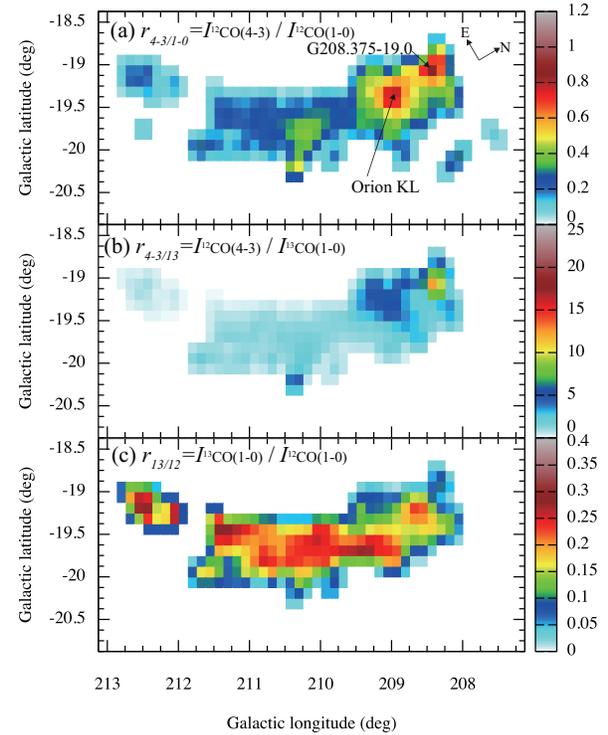}
\end{center}
\caption{Distributions of the intensity ratios of  ($a$) $r_{4-3/1-0}$, ($b$) $r_{4-3/13}$, and ($c$) $r_{13/12}$, where the data of $^{12}$CO ($J$ = 1--0) and $^{13}$CO ($J$ = 1--0) were adopted from \citet{wilson2005} and \citet{bally1987}.}
\label{ratio}
\end{figure}

Figure \ref{ratio} shows the spatial distributions of the intensity ratios.
Denser and warmer hydrogen can populate the CO molecules to higher excitation levels and produce higher rotational transition emission such as CO ($J$ = 4--3).
Thus $r_{4-3/1-0}$ would be high at the regions where the dense and/or warm gas is associated. 
$r_{4-3/13}$ and $r_{13/12}$ would be also emphasized the distributions of the warm gas and dense gas, respectively.
We examine the trend of the physical conditions of the molecular gas using the ratio maps (figure \ref{ratio}), which will be compared with the results of the RADEX calculations in the next subsection.  
There are two peaks of $r_{4-3/1-0}$ (figure \ref{ratio}a). One with $r_{4-3/1-0}=0.9\pm0.2$ at ($l$, $b$) $\approx$ (\timeform{209.0D}, \timeform{-19.375D}) corresponds to Orion KL which is the most active star forming region in the Orion-A GMC and is identified in the integrated intensity map also.
The another (we refer to it as G208.375-19.0) with $r_{4-3/1-0}=1.1\pm0.3$ is at ($l$, $b$) $\approx$ (\timeform{208.375D}, \timeform{-19.0D}), the north of Orion KL.
This peak is identified in the ratio map only. In addition, it seems that the relatively high ratio ($r_{4-3/1-0} >$ 0.5) traces the ``CO front'', which is a  region having a sharp contrast, at the eastern edge of the main component of Orion A (OMC-2/3), that is shown in a high resolution mapping of Orion A in the CO ($J$ = 1--0) line by \citet{shimajiri2011}. 
\citet{nishimura2015} also reported high $^{12}$CO ($J$=2--1)/$^{12}$CO ($J$=1--0) ratio around G208.375-19.0, while the high $r_{4-3/1-0}$ clearly shows peak value of 1.1 just at G208.375-19.0.
Their $^{13}$CO ($J$ = 2--1)/$^{12}$CO ($J$ = 2--1) map shows maxima toward G208.375-19.0, which may trace the same gas highlighted by $r_{4-3/1-0}$.

As mentioned in the introduction, some studies speculated that the diffuse gas is swept up by Orion OB associations and collided with Orion A. This may cause the heating of the gas.
We note that the star cluster NGC1977 contains a B star, HD 37018, and may be an alternative heating source of the gas around G208.375-19.0, where $r_{4-3/1-0}$ shows $1.1\pm0.3$. 
G208.375-19.0 shows high ratio of $r_{4-3/13} > 10$, while $r_{13/12}$ is moderate, $\sim0.1$.
In gaussian fit of spectra of CO ($J$ = 4--3), CO ($J$ = 1--0), $^{13}$CO ($J$ = 1--0) at G208.375-19.0, the peak velocity and the line width (FWHM) are measured to be $V_{\mathrm{LSR}} =$ 12.1 $\pm$ 0.7, 12.1 $\pm$ 0.7, and 12.2 $\pm$ 0.3 km s$^{-1}$ and $\Delta V =$ 1.6, 1.5, and 1.0 km s$^{-1}$, respectively.
This high ratio region extends like a filament that traces the CO front.
Orion KL also shows the high ratio of $r_{4-3/13} \sim 5$, but the ratio is less than that for G208.375-19.0.
Further, the $I_{^{13}\mathrm{CO (1-0)}} /I_{^{12}\mathrm{CO (1-0)}}$ map (figure \ref{ratio}c) shows the  high ratio at the southern ($l <$ \timeform{209D}) side of Orion A.
A peak of $r_{13/12}$ can be found at the inside of G208.375-19.0 near the CO front.
This would reflect that the density of the gas is increasing at the inside of the CO front.
Near L1641-N at $(l, b) \approx (\timeform{210.125D}, \timeform{-19.75D})$, $r_{4-3/1-0}=0.45$ is higher than the entire average, but the distribution of high $r_{13/12}$ (=0.23--0.3) region is wider than that of $r_{4-3/1-0}$ and has an offset from L1641-N toward the north-east direction.

\subsubsection{RADEX calculation}
We adopted RADEX \citep{vandertak2007} to estimate physical conditions of the molecular gas in the Orion A. RADEX is a non-LTE radiative transfer computer code. 
Using RADEX, we calculated the intensity ratios for given physical parameters of the molecular gas, the kinetic temperature $T_{\mathrm{kin}}$, the density of molecular hydrogen $n($H$_2)$, and the column density of $^{12}$CO per unit velocity width of the spectrum $N(^{12}$CO)$/dV$.
The ranges of individual input parameters were $T_{\mathrm{kin}}$ = 5--100 K, $n$(H$_2$)=10--10$^8$ cm$^{-3}$, and $N(^{12}$CO)$/dV$=10$^{13}$--10$^{19}$ cm$^{-2}$ (km s$^{-1}$)$^{-1}$.
We assumed the typical line width of the spectra as $dV=5.0$ km s$^{-1}$ from observations and adopted a [$^{12}$CO]/[$^{13}$CO] abundance ratio of 50 and the back ground temperature of \mbox{2.73 K.} The assumed line width would be larger than that in the peripheral region ($dV \approx 2-3$ km s$^{-1}$). However, the calculation of physical parameters, $T_{\mathrm{kin}}$, $n($H$_2)$, and $N(^{12}$CO)$/dV$ is independent of the line width as long as we use the ratio of the integrated intensities. It is to be noted that the difference of the line width should be taken into account when we estimate the column density $N(^{12}$CO) from $N(^{12}$CO)$/dV$. 

We assumed that the geometry and density structures of the molecular gas follows the standard homogenous-sphere model for the escape probability of photons in RADEX calculations.
Only H$_2$ was taken into account as the collision partner with CO molecules.
The collisional coefficients between H$_2$ molecules and $^{12}$CO/$^{13}$CO were taken from the Leiden Atomic and Molecular Database (LAMDA; \cite{schoier2005}). 
RADEX returned the integrated intensities of $^{12}$CO ($J$ = 4--3), $^{12}$CO ($J$ = 1--0), and $^{13}$CO ($J$ = 1--0) using the given physical parameters of $T_{\mathrm{kin}}$, $n$(H$_2)$, and $N(^{12}\mathrm{CO})/dV$.

The $\chi^2$ was evaluated to estimate of the physical parameters from the results of the RADEX calculation.
The $\chi^2$ is given as
\begin{eqnarray}
\chi^2=\left\{r_{4-3/1-0}-R_{4-3/1-0}(T, n, N/dV)\right\}^2\nonumber\\
+\left\{r_{4-3/13}-R_{4-3/13}(T, n, N/dV)\right\}^2\nonumber\\
+\left\{r_{13/12}-R_{13/12}(T, n, N/dV)\right\}^2,
\end{eqnarray}
where $R_{4-3/1-0}$, $R_{4-3/13}$, and $R_{13/12}$ are the calculated intensity ratios using RADEX results and are functions of $T_{\mathrm{kin}}$, $n$, and $dN/dV$.
A set of physical conditions was found as the best combination of $T_{\mathrm{kin}}$, $n_{\mathrm{H_2}}$, and $N(^{12}\mathrm{CO})/dV$ which minimized the $\chi^2$.
The average of the minimized $\chi^2$ was $3.6\times10^{-4}$, which corresponds to deviation between the observed and calculated ratios of 0.01, which is comparable with the rms of each intensity ratio.

\subsection{Spatial distribution of the physical conditions}

\begin{figure*}[tbp]
\begin{center}
\includegraphics[width=140mm]{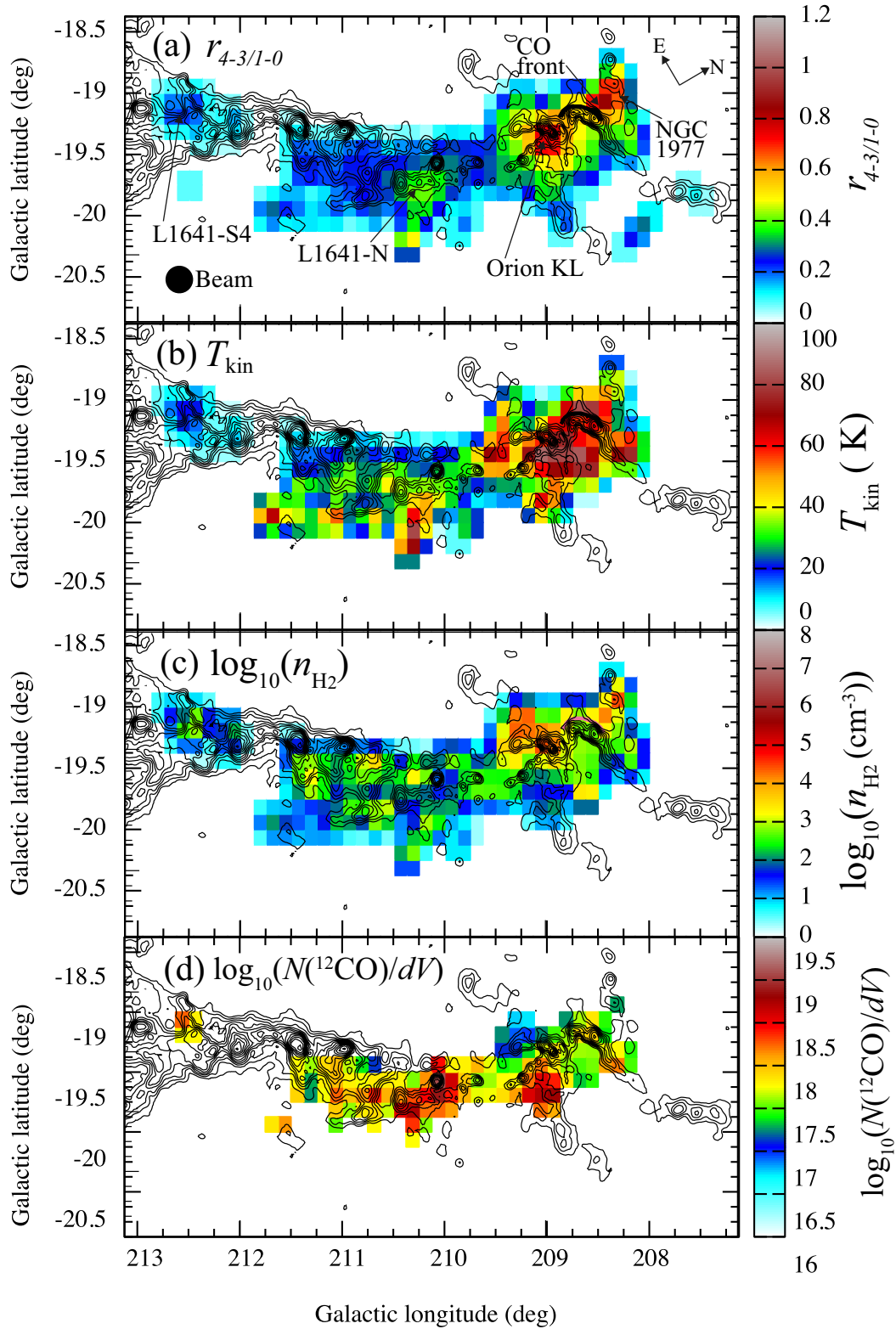}
\end{center}
\caption{Distribution of ($a$) $r_{4-3/1-0}=$ $I_{^{12}\mathrm{CO (4-3)}} /I_{^{12}\mathrm{CO (1-0)}}$, ($b$) the kinetic temperature $T_{\mathrm{kin}}$, ($c$) the volume number density of $n_{\mathrm{H_2}}$, and ($d$) $N(^{12}\mathrm{CO})/dV$ in color. The contour map shows extinction estimated from the 2MASS data \citep{dobashi2011}.}
\label{all}
\end{figure*}

Figure \ref{all} shows the distribution of the three output parameters of $T_{\mathrm{kin}}$, $n_{\mathrm{H_2}}$, and $N(^{12}\mathrm{CO})/dV$, calculated by RADEX, overlaid on an extinction map made with 2MASS data \citep{dobashi2011}.
The kinetic temperature and the density toward Orion KL are $T_{\mathrm{kin}}$ = 80--100 K and $n_{\mathrm{H_2}} \sim 10^5$ cm$^{-3}$, respectively.
These values are consistent with the results estimated so far \citep{schulz1995}.
The column density per unit velocity is $N(^{12}\mathrm{CO})/dV \sim 10^{18}$ cm$^{-2}$ (km s$^{-1}$)$^{-1}$. 
Although the number density at the west side of Orion KL decreases with the distance from Orion KL, the number density at the east side of Orion KL shows relatively high value ($\sim 10^4$ cm$^{-3}$).
In contrast, the column density $N(^{12}\mathrm{CO})/dV$ is high at the west side and low at the east side, indicating that the optical depth at the west is high and that at the east is low.

The physical parameters at G208.375-19.0 are $T_{\mathrm{kin}} \sim 80$ K and $n_{\mathrm{H_2}} \sim 10^4-10^5$ cm$^{-3}$.
The warm temperature region of $T_{\mathrm{kin}}  \sim 60$ K around G208.375-19.0 extends toward the CO front that is traced as a filamentary structure in the extinction map. 
The high density region, with $n_{\mathrm{H_2}} \sim 10^4$ cm$^{-3}$, may be localize at the northeast of the CO front near NGC1977 at ($l$, $b$) $=$ (\timeform{208.5D}, \timeform{-19.1D}) \citep{peterson2008}.
However the angular resolution ($9'.4$) of the 30-cm telescope is insufficient to determine the accurate position of G208.375-19.0.
The column density is $10^{17}$--$10^{18}$ cm$^{-2}$ (km s$^{-1}$)$^{-1}$ at G208.375-19.0, which is comparable with that of Orion KL.
There is no clear difference of the density and column density at the east and west side of G208.375-19.0.

The warm gas of  $\sim 40$ K is associated with L1641-N.
The density is not so high, $\sim 10^{3}$ cm$^{-3}$, around L1641-N, but the column density of $^{12}$CO is high, $\sim10^{18}$ cm$^{-2}$ (km s$^{-1}$)$^{-1}$.
This indicates that the optical depth in L1641-N is larger than that of the gas at the northern part of the Orion-A GMC.
In L1641-S4, the temperature and the density are less than $20$ K and $\sim 10^{3}$ cm$^{-3}$, respectively.
Several local parts of L1641-S4 show a column density of $\sim10^{18}$ cm$^{-2}$ (km s$^{-1}$)$^{-1}$.
This is less than that of L1641-N, but higher than that for the northern part of the Orion-A GMC.

The Orion-A GMC has a global gradient of the physical properties.
The temperature decreases from north ($\sim$ 100 K) to south ($\sim$ 20 K), which is basically consistent with the results derived by CO ($J$ = 2--1) and $^{12}$CO ($J$ = 1--0) emissions \citep{nishimura2015}.
Moreover the temperature is lower at the east side than that of the west side in the southern part of the Orion-A GMC.
The density distribution shows almost constant value of $\sim 10^{3}$ cm$^{-3}$, excepting for the gas associated with Orion KL and the CO front.
The column density per unit velocity of $^{12}$CO has a gradient along the north-south direction.
The column density increases from the northern part  [$\sim10^{17.5}$ cm$^{-2}$ (km s$^{-1}$)$^{-1}$] to the southern part [$\sim10^{18-19}$ cm$^{-2}$ (km s$^{-1}$)$^{-1}$)] where the L1641-N region is included.
Medians of $T_{\mathrm{kin}}$, $n_{\mathrm{H_2}}$, and $N(^{12}\mathrm{CO})/dV$ in the entire observed region of figure \ref{all} are 34 K, $3.5\times10^{2}$ cm$^{-3}$, and $5.3\times10^{17}$ cm$^{-2}$ (km s$^{-1})^{-1}$, respectively. This median of the density is consistent with that reported by \citet{sakamoto1994}.

\subsection{Physical property and triggered star formation}
G208.375-19.0 contains warm and dense gas and associates with the CO front.    
Recently, several large telescopes have started to survey the Orion-A GMC with higher angular resolution that mainly focused on the detailed structure of the GMC such as filaments, clumps, cores and outflows in particular parts of the cloud.
\citet{shimajiri2011}  clearly showed the diffuse CO component is interacting with the CO front and is being accumulated on to the surface of the main cloud component at the CO front (Figure 14 in \cite{shimajiri2011}). 
Some authors suggested that the Orion OB 1a or 1b group, located at $\sim$100 pc east from the Orion-A GMC affected Orion-A by its strong UV emission ($G_0=10^{4-5}$) and stellar wind \citep{tielens1985, bally1987, wilson2005}.
The radiation pressure of these OB stars may sweep up the pre-existing diffuse gas toward Orion A and trigger the next star formation episode in the cloud (e.g., \cite{elmegreen1977}).
Similarly, G208.375-19.0 would be interpreted as the result of the heating and accumulating of the diffuse material affected by the Orion OB association.
G208.375-19.0 is also close to NGC1977 and may be affected by B stars in NGC1977, alternatively.
Unfortunately, the low angular resolution of the 30-cm telescope cannot determine the spatial relation of G208.375-19.0 and the CO front or NGC1977. 
Observations with higher resolution are needed to confirm the origin of the high ratio at G208.375-19.0. 

\section{Conclusion}
The $^{12}$CO ($J$ = 4--3) line emission in the Orion-A GMC has been mapped with a $9'.4 \pm 0'.4$ resolution using the Tsukuba 30-cm telescope.
We investigated the physical properties of the molecular gas over the clouds.
The conclusions are summarized as follows:  

\begin{enumerate}
\item The emission of CO ($J$ = 4--3) extends over 3 deg$^2$ with $T_{\mathrm{mb}} > 2.0$ K, covering the main component of the molecular gas in the Orion-A GMC.
The most intense emission of $T_{\mathrm{mb}}=27$ K was detected toward the Orion KL region.  
The other star forming region, L1641-N, also has a peak in the integrated intensity map.

\item The velocity gradient from south to north of the main component of the GMC is observed in the same way for the $^{12}$CO ($J$ = 1--0) and $^{13}$CO ($J$ = 1--0) emission lines.
A velocity jump at $l \approx 208.5^{\circ}$ region is also traced in the longitude-velocity map. 

\item The integrated intensity ratios between CO ($J$ = 4--3) and CO ($J$ = 1--0) is $r_{4-3/1-0} \sim 0.2$ in the southern region and 0.4--0.8 at the star forming regions of the cloud.
Orion KL and a position near G208.375-19.0 on the CO front shows maxima in the map of the ratio, $0.9\pm0.2$ and $1.1\pm0.3$, respectively. 
G208.375-19.0 is clearly manifested for the first time in this $r_{4-3/1-0}$ ratio map.

\item We derived physical properties of the molecular gas using non-LTE analysis. 
The kinetic temperature of the star forming regions, Orion KL and L1641-N, are $T_{\mathrm{kin}}$ = 80--100 K and 40 K, respectively.
These results are consistent with the values estimated by previous observations in lower transition lines of CO and its isotopologue.
The results also indicate a gradient of the temperature that ranges within $T_{\mathrm{kin}}$=20 K in the south to 80 K in the north.

\item The non-LTE analysis shows that the kinetic temperature and the density are increasing near the CO front including G208.375-19.0. The gas associated with the edge of the northeast part of the cloud including the CO front is warm ($T_{\mathrm{kin}}\sim60$ K), dense ($n_{\mathrm{H_2}}\sim10^4$ cm$^{-3}$), and optically thin. This indicates the global compression of the gas in the northern part of Orion A. 
This may be caused by the heating by UV emission from OB stars or compressing of the gas by an accumulation of the diffuse gas to the CO front that relates with the triggered star formation. However, observations of the CO line from high- and middle-excited molecules with high angular resolution are needed to determine the location of the origin of this warm and dense gas.
\end{enumerate}

\section*{Acknowledgement}
We are grateful to M. Ogino, Y. Ishizaki, M. Nakano, H. Okura, Y. Terabe, Y. Nihonmatsu, Y. Koide, D. Salak, K. Saito, K. Doihata, and N. Reyes for helping with the development of the 30-cm telescope and its operation. 
We acknowledge the Corporaci\'on Nacional Forestal (CONAF) at Putre and people at Parinacota in Chile for the support during the operation of the telescope. 
The International Foundation High Altitude Research Stations Jungfraujoch and Gornergrat (HFSJG) made it possible for us to carry out experiment at the station at Jungfraujoch in Switzerland. This work was supported by JSPS KAKENHI Grant Number 19204016, 22244011, and 26247019 and research grants of the Toray Science Foundation and the Mitsubishi Foundation. LB and RF gratefully acknowledge support by CONICYT Grant PFB-06 and FONDECYT project 1120195.


\begin{thebibliography}{99}
\expandafter\ifx\csname natexlab\endcsname\relax\def\natexlab#1{#1}\fi

\bibitem[{Bally {et~al.}(1987)Bally, Lanber, Stark, \& Wilson}]{bally1987}
Bally, J., Lanber, W. D., Stark, A. A., \& Wilson, R. W. 1987, \apj, 312, L45

\bibitem[{Bronfman {et~al.}(1988)Bronfman, Cohen, Alvarez, May, \&
  Thaddeus}]{bronfman1988}
Bronfman, L., Cohen, R. S., Alvarez, H., May, J., \& Thaddeus, P. 1988, \apj, 324, 248

\bibitem[{Brown {et~al.}(1994)Brown, de~Geus, \& de~Zeeuw}]{brown1994}
Brown, A. G. A., de~Geus, E. J., \& de~Zeeuw, P. T. 1994, \aap, 289, 101

\bibitem[{Dame {et~al.}(1987)Dame, Ungerechts, Cohen, De~Geus, Grenier, May, Murphy, Nyman, \& Thaddeus}]{dame1987}
Dame, T. M.,  {et~al.} 1987, \apj,  322, 706

\bibitem[{Dame {et~al.}(2001)Dame, Hartmann, \& Thaddeus}]{dame2001}
Dame, T. M., Hartmann, D., \& Thaddeus, P. 2001, \apj,  547, 792

\bibitem[{Dobashi {et~al.}(1994)}]{dobashi1994}
Dobashi, K., Bernard, J.-P., Yonekura, Y., \& Fukui, Y. 1994, \apjs, 95, 419

\bibitem[{Dobashi(2011)}]{dobashi2011}
Dobashi, K. 2011, \pasj, 63, S1

\bibitem[{Elmegreen \& Lada(1977)}]{elmegreen1977}
Elmegreen, B. G., \& Lada, C. J. 1977, \apj, 214,  725

\bibitem[{Fukui(1989)}]{fukui1989}
Fukui, Y. 1989, ESO Conf. and Workshop Proc., 33, 95

\bibitem[{Hirota {et~al.}(2007)Hirota, Ando, Bushimata, Choi, Honma, Imai, Iwadate, Jike, Kameno, Kameya, {et~al.}}]{hirota2007}
Hirota, T., {et~al.} 2007, \pasj, 59, 897

\bibitem[{Ikeda {et~al.}(1999)Ikeda, Maezawa, Ito, Saito, Sekimoto, Yamamoto, Tatematsu, Arikawa, Aso, Noguchi, {et~al.}}]{ikeda1999}
Ikeda, M., {et~al.} 1999, \apj, 527, L59
  
\bibitem[{Ikeda {et~al.}(2002)Ikeda, Oka, Tatematsu, Sekimoto, \& Yamamoto}]{ikeda2002}
Ikeda, M., Oka, T., Tatematsu, K., Sekimoto, Y., \& Yamamoto, S. 2002, \apjs, 139, 467
  
\bibitem[{Ishii {et~al.}(2013)Ishii, Seta, Nakai, Miyamoto, Nagai, Arai, Maezawa, Nagasaki, Miyagawa, Motoyama, {et~al.}}]{ishii2013}
Ishii, S., {et~al.} 2013, IEEE Transaction on Terahertz Science and  Technology, 3, 15

\bibitem[{Johnstone \& Bally(1999)}]{johnstone1999}
Johnstone, D., \& Bally, J. 1999, \apj, 510, L49

\bibitem[{Johnstone \& Bally(2006)}]{johnstone2006}
Johnstone, D., \& Bally, J. 2006, \apj, 653, 383

\bibitem[{Kim {et~al.}(2008)Kutner, Tucker, Chin, \& Thaddeus}]{kim2008}
Kim, M. K., {et~al.} 2008, \pasj, 60, 991
  
\bibitem[{Kutner {et~al.}(1977)Kutner, Tucker, Chin, \& Thaddeus}]{kutner1977}
Kutner, M. L., Tucker, K. D., Chin, G., \& Thaddeus, P. 1977, \apj, 215, 521

\bibitem[{Linsky(1973)}]{linsky1973}
Linsky, J.L. 1973, \solphys, 28, 409

\bibitem[{Maddalena {et~al.}(1986)Maddalena, Morris, Moscowitz, \&  Thaddeus}]{maddalena1986}
Maddalena, R. J., Morris, M., Moscowitz, J., \& Thaddeus, P. 1986, \apj, 303, 375

\bibitem[{Manabe {et~al.}(2003)Manabe, Inatani, Murk, Wylde, Seta, \& Martin}]{manabe2003}
Manabe, T., Inatani, J., Murk, A., Wylde, R., Seta, M., \& Martin, D. H. 2003, IEEE Transactions on Microwave Theory and Techniques, 51, 6

\bibitem[{Menten {et~al.}(2007)}]{menten2007}
Menten, K. M., Reid, M. J., Forbrich, J., \& Brunthaler, A. 2007, \aap, 474, 515

\bibitem[{Mizuno {et~al.}(1995)}]{mizuno1995}
Mizuno, A., Onishi, T., Yonekura, Y., Nagahama, T., Ogawa, H., \& Fukui, Y. 1994, \apjl, 445, L161

\bibitem[{Nagahama {et~al.}(1998)Nagahama, Mizuno, Ogawa, \& Fukui}]{nagahama1998}
Nagahama, T., Mizuno, A., Ogawa, H., \& Fukui, Y. 1998, \apj, 116, 336

\bibitem[{Nakajima {et~al.}(2007)Nakajima, Kaiden, Korogi, Kimura, Yonekura, Ogawa, Nishiura, Dobashi, Handa, Kohno, {et~al.}}]{nakajima2007a}
Nakajima, T., {et~al.} 2007, \pasj, 59, 1005

\bibitem[{Nakamura {et~al.}(2012)Nakamura, Miura, Kitamura, Shimajiri, Kawabe, Akashi, Ikeda, Tsukagoshi, Momose, Nishi, {et~al.}}]{nakamura2012}
Nakamura, F., {et~al.} 2012, \apj, 746, 25

\bibitem[{Nishimura {et~al.}(2015)Nishimura, Tokuda, Kimura, Muraoka, Maezawa, Ogawa, Dobashi, Shimoikura, Mizuno, Fukui, {et~al.}}]{nishimura2015}
Nishimura, A., {et~al.} 2015, \apjs, 216, 1

\bibitem[{Oka {et~al.}(1998)}]{oka1998}
Oka, T., Hasegawa, T., Hayashi, M., Handa, T., \& Sakamoto, S. 1998, \apj, 493, 730

\bibitem[{Penzias \& Burrus(1973)}]{penzias1973}
Penzias A. A., \& Burrus C. A. 1973, \araa, 11, 51

\bibitem[{Peterson \& Megeath(2008)}]{peterson2008}
Peterson D.~E., \& Megeath S.~T., 2008, Handbook of Star Forming Regions, Volume I, 590

\bibitem[{Sakamoto {et~al.}(1994) Sakamoto, Hayashi, Hasegawa, Handa, \& Oka}]{sakamoto1994}
Sakamoto, S., Hayashi, M., Hasegawa, T., Handa, T., \& Oka, T. 1994, \apj, 425, 641

\bibitem[{Sch{\"o}ier, {et~al.}(2005) Sch{\"o}ier, van der Tak, van Dishoeck, \& Black}]{schoier2005}
Sch{\"o}ier, F.~L., van der Tak, F.~F.~S., van Dishoeck, E.~F., \& Black, J.~H. 2005, \aap, 432, 369

\bibitem[{Schulz {et~al.}(1995) Schulz, Henkel, Beckmann, Kaseman, Schneider, Nyman, Persson, Gunnarsson, \& Delgado}]{schulz1995}
Schulz, A., Henkel, C., Beckmann, U., Kaseman, C., Schneider, G., Nyman, L. A., Persson, G., Gunnarsson, L. G., \& Delgado, G., \aap, 295, 183

\bibitem[{Seta {et~al.}(2012) Seta, Nakai, Ishii, Nagai, Miyamoto, Ichikawa, Takato, \& Motoyama}]{seta2012}
Seta, M., Nakai, N., Ishii, S., Nagai, M., Miyamoto,  Y., Ichikawa, T., Takato, N., \& Motoyama, H. 2012, Proc. IAU, 8, 251

\bibitem[{Shimajiri {et~al.}(2011) Shimajiri, Kawabe, Takakuwa, Saito, Tsukagoshi, Momose, Ikeda, Akiyama, Austermann, Ezawa.  {et~al.}}]{shimajiri2011}
Shimajiri, Y., {et~al.} 2011, \pasj, 63, 105

\bibitem[{Tielens \& Hollenbach(1985)}]{tielens1985}
Tielens, A. G. G. M., \& Hollenbach, D. 1985, \apj, 291, 747

\bibitem[{Ulich \& Haas(1976)}]{ulich1976}
Ulich, B. L., \& Haas, R. W. 1976, \apjs, 30, 247

\bibitem[{Van der Tak {et~al.}(2007)}]{vandertak2007}
Van der Tak, F. F. S., Black, J. H., Sch{\"o}ier, F. L., Jansen, D. J., van Dishoeck, E. F. 2007, \aap, 468, 627

\bibitem[{Wilson {et~al.}(2005)Wilson, Dame, Masheder, \& Thaddeus}]{wilson2005}
Wilson, B. A., Dame, T. M., Masheder, M. R. W., \& Thaddeus, P. 2005, \aap, 430, 523

\bibitem[{Yoda {et~al.}(2010)Yoda, Handa, Kohno, Nakajima, Kaiden, Yonekura, Ogawa, Morino, \& Dobashi}]{yoda2010}
Yoda, T., {et~al.}  2010, \pasj, 62, 1277

\end{thebibliography}
\end{document}